\documentclass[fleqn,usenatbib]{mnras}

\usepackage{newtxtext,newtxmath}
\usepackage[T1]{fontenc}

\DeclareRobustCommand{\VAN}[3]{#2}
\let\VANthebibliography\thebibliography
\def\thebibliography{\DeclareRobustCommand{\VAN}[3]{##3}\VANthebibliography}

\usepackage{graphicx}	% Including figure files
\usepackage{amsmath}	% Advanced maths commands

\title{New Insights on $\nu$-DM Interactions}

\author[]{
Philippe Brax,$^{1,2}$%\thanks{E-mail: publications@ras.ac.uk (KTS)}
Carsten van de Bruck,$^{3}$
Eleonora Di Valentino,$^{3}$
William Giar\'e,$^3$ 
and Sebastian Trojanowski$^{4.5}$
\\
$^{1}$Institut de Physique Théorique, Université Paris-Saclay,
CEA, CNRS, F-91191 Gif-sur-Yvette Cedex, France\\
$^{2}$CERN, Theoretical Physics Department, Geneva, Switzerland\\
$^{3}$
School of Mathematics and Statistics, University of Sheffield,
Hounsfield Road, Sheffield S3 7RH, United Kingdom\\
$^{4}$
Astrocent, Nicolaus Copernicus Astronomical Center Polish Academy of Sciences, ul. Rektorska 4, 00-614, Warsaw, Poland\\
$^{5}$
National Centre for Nuclear Research, ul. Pasteura 7, 02-093 Warsaw, Poland
}

\date{Accepted XXX. Received YYY; in original form ZZZ}

\pubyear{2015}

\begin{document}
\label{firstpage}
\pagerange{\pageref{firstpage}--\pageref{lastpage}}
\maketitle

\begin{abstract}
We revisit the possibility of using cosmological observations to constrain models that involve interactions between neutrinos and dark matter. We show that small-scale measurements of the cosmic microwave background with a few per cent accuracy are critical to uncover unique signatures from models with tiny couplings that would require a much higher sensitivity at lower multipoles, such as those probed by the Planck satellite. We analyze the high-multipole data released by the Atacama Cosmology Telescope, both independently and in combination with Planck and Baryon Acoustic Oscillation measurements, finding a compelling preference for a non-vanishing coupling, $\log_{10}u_{\nu \textrm{DM}}=-5.20^{+1.2}_{-0.74}$ at 68\% CL. This aligns with other CMB-independent probes, such as Lyman-$\alpha$. We illustrate how this coupling could be accounted for in the presence of dark matter interactions with a sterile neutrino.
\end{abstract}


%%%%%%%%%%%%%%%%%%%%%%%%%%%%%%%%%%%%%%%%%%%%%%%%%%

%%%%%%%%%%%%%%%%% BODY OF PAPER %%%%%%%%%%%%%%%%%%

\section{Introduction}
Precision measurements of Cosmic Microwave Background (CMB) radiation~(\cite{Planck:2018vyg, ACT:2020frw, SPT-3G:2022hvq}) have substantially furthered our understanding of dark matter (DM) by offering a convincing, albeit indirect,  supporting evidence for its existence and precise constraints on its properties. Nevertheless, despite these advances, DM is still elusive, as confirmed by a variety of  unsuccessful experiments, including direct searches and astrophysical observations.

The enigmatic nature of DM can be attributed to its poorly understood interactions with other particles: apart from gravitational interactions, its fundamental couplings to the Standard Model remain unknown and debated. Building on this unresolved uncertainty surrounding the interaction strengths of DM with other particles, a fascinating and persistent idea is the possibility of a coupling between DM and neutrinos through an as-yet-undiscovered interaction channel. The literature offers a wide range of possible forms of the cross section governing such interactions, with significant implications for various observables spanning from cosmology to astrophysics and accelerator-based searches~(\cite{Palomares-Ruiz:2007trf,Serra:2009uu,Shoemaker:2013tda,Wilkinson:2013kia,Wilkinson:2014ksa,Olivares-DelCampo:2017feq,Bertoni:2014mva,Batell:2017rol,Batell:2017cmf,DiValentino:2017oaw,Escudero:2018thh,Kolb:1987qy,Shoemaker:2015qul,deSalas:2016svi,Pandey:2018wvh,Kelly:2019wow,Blennow:2019fhy,Choi:2019ixb,Kelly:2021mcd}).
In this work, we revisit the possibility of using CMB observations to constrain models that involve interactions between neutrinos and dark matter ($\nu$DM) described in terms of a single parameter
\begin{equation}
	u_{\nu \textrm{DM}} = \frac{\sigma_{\nu\textrm{DM}}}{\sigma_{\textrm{T}}}\,\left(\frac{m_{\textrm{DM}}}{100~\textrm{GeV}}\right)^{-1},
	\label{eq:defu}
\end{equation}
where $\sigma_{\nu\textrm{DM}}$ and $\sigma_{\textrm{T}}$ are the  $\nu$DM and Thomson scattering cross sections and $m_{\rm DM}$ is the mass of the dark matter particle, respectively. The impact of such an interaction on the CMB angular power spectra and the late-time matter power spectrum can be significant, depending on its strength. Therefore, extensive studies have been conducted to understand the cosmological implications of these effects and constraints from current CMB and large-scale structure observations, as well as forecasts for next-generation surveys~(\cite{Escudero:2015yka}), are available in the literature. 

\begin{figure}
    \centering
    \includegraphics[width=0.46\textwidth]{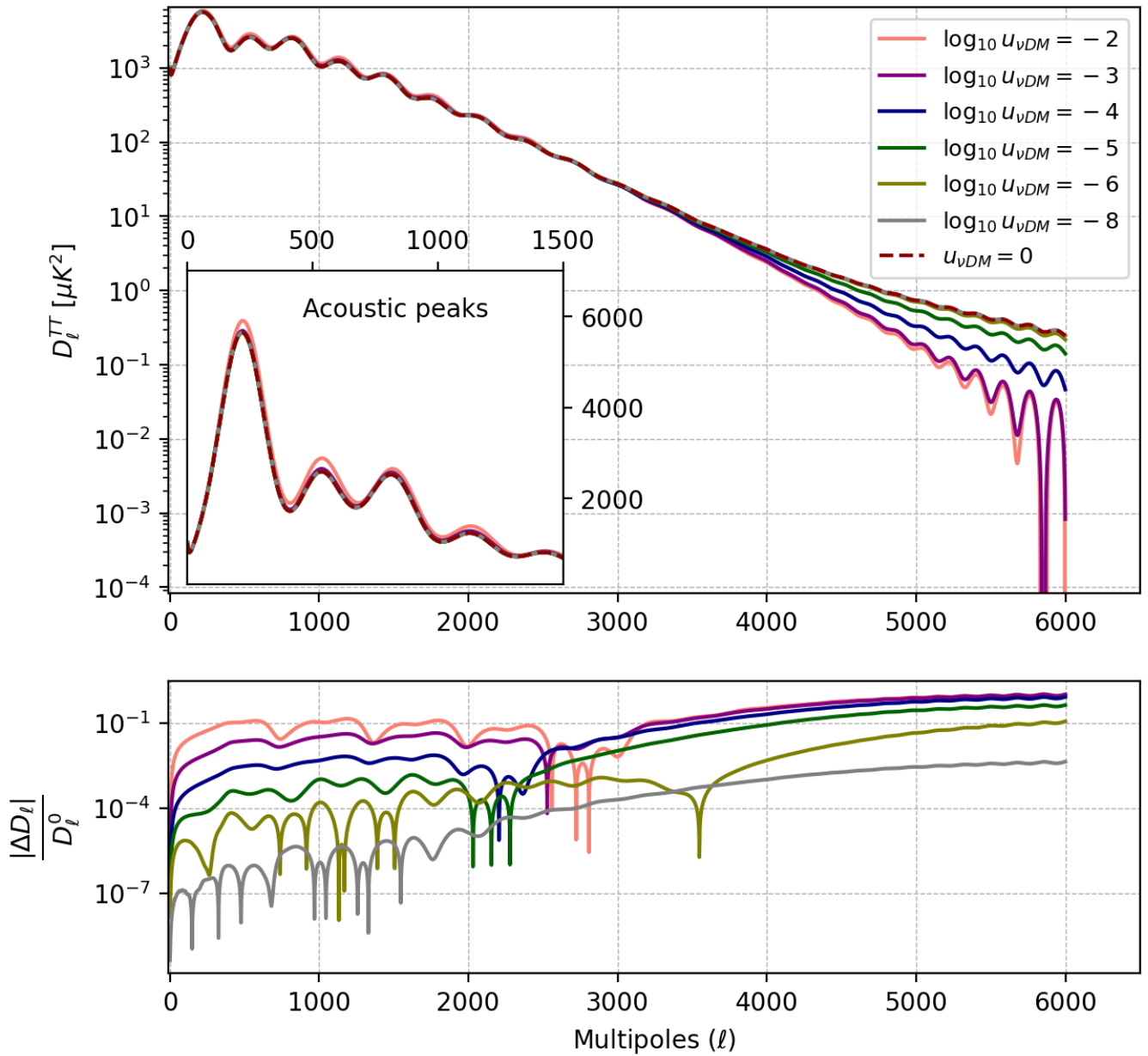}
    \caption{\small The top panel displays the theoretical $D_{\ell}^{TT}$, while the percentage difference $|\Delta D_{\ell}| / D_{\ell}^{0}$ with respect to the non interacting case ($D_{\ell}^{0}$) for different coupling values is shown in the bottom panel. The figure highlights that feeble interactions can result in undetectable changes in the Planck's probed multipole range, but can produce substantial differences on smaller scales (i.e., higher multipoles) like those measured by ACT.}
    \label{fig:theory}
\end{figure}

The state-of-the-art cosmological analyses on $\nu$DM interactions arise primarily from the CMB data released by the Planck collaboration, which provides precise measurements of the angular power spectra of temperature and polarization anistropies in the multipole range $2\lesssim \ell \lesssim 2500$.  Assuming a temperature-independent $\nu$DM cross section $\sigma_{\nu\textrm{DM}}\sim T^0$, constraints on the interaction strength can be derived, typically resulting into upper limits $u_{\nu \textrm{DM}} \leq (4.5-9.0) \times 10^{-5}$ at 95\% CL~(\cite{DiValentino:2017oaw,Mosbech:2020ahp,Paul:2021ewd}).  As clearly shown in the bottom panel of~\autoref{fig:theory}, these bounds reflect the limited (albeit remarkable) sensitivity reachable by CMB observations. Indeed, on the scales probed by experiments similar to Planck, values $u_{\nu \textrm{DM}}\lesssim 10^{-5}$ would produce corrections smaller than one part in $10^{5}$ when compared to the non-interacting case. This implies that any differences between the two cases would essentially be undetectable as it would require a precision well beyond the current accuracy of data.

However, the key observation underlying our study is that small couplings have a more significant impact on smaller scales (higher multipoles), where differences can reach a few percents when compared to the non-interacting case, as illustrated in~\autoref{fig:theory}. Therefore experiments with high precision in the damping tail at $\ell \gtrsim 3000$ provide a unique opportunity to gain novel insight into models that would otherwise be indistinguishable at lower multipoles. This holds true for both the next-generation of CMB experiments and recent measurements of the CMB angular power spectra released by ground-based telescopes. In fact, by probing higher multipoles than the Planck satellite, these measurements can provide valuable complementary information that can improve the sensitivity of current results and contribute to the study of $\nu\textrm{DM}$ interactions.

\section{Analysis}

\label{sec:Analysis}

Based on previous considerations, we extend the state-of-the-art analyses on neutrino dark matter interactions, investigating the impact of recent CMB measurements obtained from ground-based telescopes. Our analysis focuses specifically on the Atacama Cosmology Telescope (ACT) temperature and polarization DR4 likelihood~\citep{ACT:2020frw}, which explores higher multipoles ($600\lesssim \ell \lesssim 4500$) compared to the full Planck 2018 likelihood ($2\lesssim \ell \lesssim 2500$)~\cite{Planck:2019nip,Planck:2018vyg,Planck:2018nkj}. This produces precise data on small scales where the effects of small couplings start to become comparable with the observational constraining power. Additionally, alongside CMB observations, we take into account measurements of Baryon Acoustic Oscillations (BAO) and Redshift Space Distortions (RSD) from the Baryon Oscillation Spectroscopic Survey (BOSS DR12)~\citep{BOSS:2012dmf}.

To parameterize our cosmological model, we employ a common approximation in the literature, i.e. treating neutrinos as massless and ultra-relativistic in the early universe. This simplifies calculations for scenarios involving interactions with DM. In addition, we examine the interplay between neutrinos and the entire fraction of energy-density associated with DM, with a specific focus on a temperature-independent cross-section. By doing so, we only need one extra parameter in addition to the usual six $\Lambda$CDM parameters, which is the logarithm of the coupling parameter $\log_{10}u_{\nu\rm{DM}}$, as defined in \autoref{eq:defu}. To compute the cosmological model and study the effects of $\nu\textrm{DM}$ interactions, we make use of a modified version of the Cosmic Linear Anisotropy Solving System code \texttt{CLASS}\footnote{A publicly available version can be found at \url{https://github.com/MarkMos/CLASS_nu-DM}, see also Refs.~(\cite{Stadler:2019dii,Mosbech:2020ahp}).}~(\cite{Blas:2011rf}). We explore the posterior distributions of our parameter space by exploiting the publicly available code \texttt{COBAYA}~(\cite{Torrado:2020xyz}) and the Monte Carlo Markov Chain (MCMC) sampler developed for \texttt{CosmoMC}~(\cite{Lewis:2002ah}).

\begin{figure}
    \centering
    \includegraphics[width=\columnwidth]{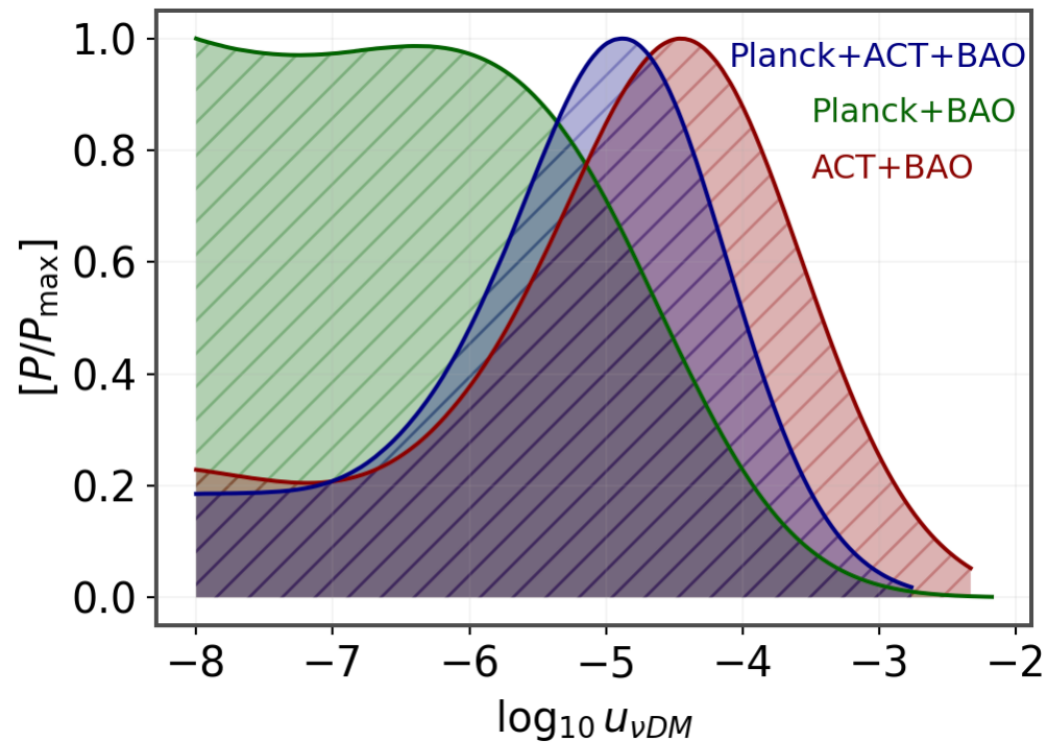}
    \caption{\small Posterior probability distribution functions for the coupling $\log_{10} u_{\nu\text{DM}}$ resulting from different combinations of CMB and BAO+RSD measurements.}
    \label{fig:data}
\end{figure}

Firstly, by considering the full temperature and polarization Planck likelihood in the multipole range of $2\lesssim \ell \lesssim 2500$, in combination with BAO and RSD measurements, we are able to replicate the results previously discussed in the literature  yielding an upper bound of $\log_{10}u_{\nu \textrm{DM}}< -4.39$ at a 95\% CL. \autoref{fig:data} displays (in green) the posterior distribution function of $\log_{10}u_{\nu \textrm{DM}}$ for this combination of data. As illustrated in the figure, below a certain threshold of $u_{\nu \textrm{DM}}\lesssim 10^{-5}$, all the models become indistinguishable, leading to a flat posterior distribution for smaller values.

In order to investigate the impact of small-scale CMB observations, we first consider the ACT data in combination with BAO and RSD measurements. In \autoref{fig:data}, we display (in red) the posterior distribution function for this case. It is interesting to note that, as evident from the figure, the posterior distribution function for this combination of data shows a clear preference for a non-zero coupling. This preference is translated into a 68\% CL result $\log_{10}u_{\nu \textrm{DM}}=-4.86^{+1.5}_{-0.83}$. Although this indication is not supported by the Planck data, it is crucial to observe that the two datasets are not in tension regarding the predicted value for this parameter. The ACT’s indication for a non-zero coupling can be explained by the larger effects of couplings of the order of $u_{\nu \textrm{DM}}\sim 10^{-6} - 10^{-4}$ in the multipole range probed by this experiment, see \autoref{fig:theory}. Therefore, while the effects of such a tiny coupling may not be detectable at the scales probed by Planck, they may be easier to unveil at the scales measured by ACT. It is also important to note that for smaller values ($u_{\nu \textrm{DM}}\lesssim 10^{-6}$), the effects of a possible interaction between neutrinos and Dark Matter, although remaining some orders of magnitude larger than the scales probed by Planck, become too small to be distinguishable from the non-interacting case, even on multipoles probed by ACT (see \autoref{fig:theory}). Consequently, the posterior distribution function also becomes flat (see \autoref{fig:data}). As a result of this effect, we lose the indication for a non-zero coupling at a 95\% confidence level, obtaining only an upper limit of $\log_{10}u_{\nu \textrm{DM}}<-3.70$. However, this loss of evidence is related to the currently limited precision of data rather than a real preference for zero coupling values.

To validate further our argument that the preference for a non-vanishing $\nu$DM interaction comes from the high-ell ACT multipoles, we combine Planck data between $2\lesssim \ell \lesssim 650$ with the small-scale DR4 ACT likelihood, along with BAO and RSD measurements\footnote{Note that the cut made to the Planck data is necessary to avoid including the region where the two experiments overlap, which would result in double counting of the same sky in the absence of a covariance matrix~\cite{ACT:2020gnv}.}. We show the posterior distribution function of this case in \autoref{fig:data} (blue line). As evident, the preference for a non-zero interaction rate is maintained by combining the two most precise CMB experiments, so that we obtain a robust indication $\log_{10}u_{\nu \textrm{DM}}=-5.20^{+1.2}_{-0.74}$ at the 68\% CL. It is also important to note that including low-$\ell$ Planck data narrows the peak amplitude of the posterior distribution around its central value, leading to a stronger indication for an interaction between the two species. This improvement is due to the fact that Planck data provide information around the first acoustic peaks, which are not probed by ACT. Since values of $u_{\nu \textrm{DM}} \gtrsim 10^{-4}$ substantially increase the amplitude of the first acoustic peaks (see \autoref{fig:theory}), including precise measurements at lower multipoles improves the constraints in this region, leading to the observed shift in $\log_{10}u_{\nu \textrm{DM}}$. This improvement also helps to isolate the impact of $\nu\textrm{DM}$ on the ACT data by breaking the degeneracy with other cosmological parameters and shifting their values back close to $\Lambda\textrm{CDM}$ preferred values obtained with the full Planck dataset.
Nonetheless, it is important to emphasize that for interaction strengths below a certain threshold ($u_{\nu \textrm{DM}}\lesssim 10^{-6}$), the same considerations as mentioned in the ACT-only case apply to this scenario where both ACT and Planck data are combined. In other words, the impact of such weak couplings on the CMB angular spectra becomes too small compared to the data accuracy in both the Planck and ACT multipole ranges. As a consequence, all models become indistinguishable, and the posterior distribution function becomes flat, as shown in \autoref{fig:data}. This behavior of the posterior distribution function prevents us from obtaining a two-sigma constraint. Thus, at the 95\% confidence level, we can only derive an upper limit of $\log_{10}u_{\nu \textrm{DM}}<-4.17$.

\section*{Example\label{sec:DMnuinteractions}}

Given the preference in the cosmological data that we find towards non-diminishing DM-neutrino interactions, it is useful to  consider briefly the  implications of our findings for a sample specific scenarios of beyond the Standard Model (BSM) neutrino interactions. We note that for $m_{\rm DM} \sim 1~\textrm{GeV}$ the $1\sigma$ ranges of the $\sigma_{\textrm{DM}-\nu}$ cross section obtained in our analysis  correspond to values of the order of at least one nano-barn, while being even larger for heavier DM species. As a result, it is challenging to  couple directly DM to the $SU(2)_L$ lepton doublet in the SM with such a large cross-sections without violating stringent {DM direct detection bounds from electron scatterings, cf. Ref.~(\cite{Akerib:2022ort}) for recent review. Large couplings between DM and charged leptons are further constrained by missing energy searches at Large Electron–Positron
Collider (LEP) and indirect detection searches for DM annihilations into charged leptons~(\cite{Shoemaker:2013tda,Blennow:2019fhy}).} 

This can be circumvented in models employing a mixing between active and sterile neutrinos together with a  coupling of the sterile neutrinos to the the DM species~(\cite{Bertoni:2014mva,Batell:2017rol,Batell:2017cmf}).\footnote{While the strongest experimental bounds are associated with DM couplings to electrons and quarks, they could also be avoided in models employing light DM particles with flavor non-universal couplings to muons or tau leptons and to respective neutrinos, e.g., the $U(1)_{L_\mu-L_\tau}$ gauge boson portal to DM. We leave a detailed investigation of such scenarios for future studies.} For instance, a new Dirac fermion $N$ could interact with the SM via Yukawa-like couplings $\mathcal{L} \supset -\lambda\,(\bar{L}\,\hat{H})\,N_R$, where $L$ is the SM lepton doublet and $H$ is the Higgs field. This gives rise to a mixing between the active and sterile neutrinos  after electroweak symmetry breaking. The coupling of DM to $N$ is given by $\mathcal{L} \supset -\,\phi\,\bar{\chi}\,(y_L\,N_L+y_R\,N_R) + \textrm{h.c.}$, where additional fermionic $\chi$ and scalar $\phi$ SM-singlet fields have been introduced. Both of them can play the role of DM after imposing additional $U(1)_d$ symmetry, depending on which one is the lightest of the BSM species. The heavier sterile neutrino dominantly decays into the dark states, $N\to \chi\phi$, therefore alleviating constraints from visibly decaying heavy neutral leptons~(\cite{Abdullahi:2022jlv,Batell:2022xau}). 

In the mass-degenerate regime in the dark sector, $m_{\textrm{DM}} \equiv m_\chi \simeq m_\phi$, the $\chi$ DM elastic scatterings off neutrinos mediated by $\phi$ are characterized by an effectively temperature-independent cross section,
\begin{equation}
\sigma_{\textrm{DM}-\nu} \simeq 10^{-34}\,\left(\frac{g}{0.01}\right)^4\,\left(\frac{20~\textrm{MeV}}{m_{\textrm{DM}}}\right)^2\,\textrm{cm}^2,
\end{equation}
where $g = y_L\,(|U_{e4}|^2+|U_{\mu 4}|^2+|U_{\tau 4}|^2)^{1/2}$ and $U_{\ell 4}$ is the mixing angle between the sterile and active neutrino of a given flavor $\ell$. In the following, we will assume  that the dominant  mixing is with the tau neutrino, while we set other mixing angles to zero. We also take $m_N = 10\,m_{\textrm{DM}}$. In \autoref{fig:DMnuinteractionsgeneral}, we illustrate a region in the parameter space of this BSM model in the $(m_{\textrm{DM}},g)$ plane, in which one can simultaneously fit the cosmological bounds and avoid other constraints. {At the  top of \autoref{fig:DMnuinteractionsgeneral}, we show  the grey-shaded region corresponding to an upper bound on the coupling constant $g$ above which one predicts too large active-neutrino mixing angles for $y_L = 1$. The leading constraints on $U_{\tau 4}$, in this case, arise from atmospheric neutrino oscillation analyses, leptonic and semi-leptonic tau decays, and measurements of the lepton flavour universality in $B$ meson decays,  see ~(\cite{Batell:2017cmf,Cvetic:2017gkt,BaBar:2022cqj}). Light DM species that thermalize in the early Universe due to their interactions with neutrinos are subject to additional bounds from their possible contribution to the number of relativistic degrees of freedom, $N_{\textrm{eff}}$, which excludes DM mass below $\mathcal{O}(10~\textrm{MeV})$~(\cite{Boehm:2013jpa}). We note that bounds from  {heavy neutral lepton} decays during the Big Bang Nucleosynthesis (BBN) epoch can be avoided as $N$ decays preferably in the dark sector in this scenario.

\begin{figure}
    \centering
    \includegraphics[width=\columnwidth]{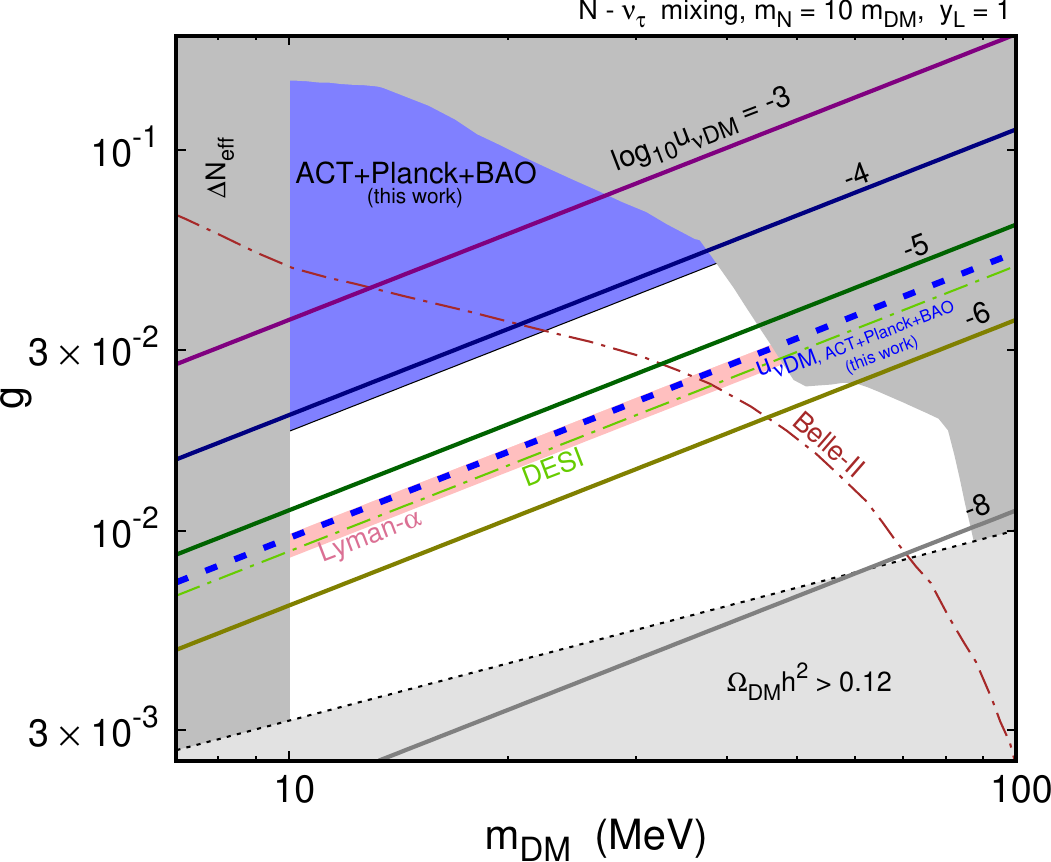}
 \caption{\small The parameter space of the neutrino portal DM model shown in the $(m_{\textrm{DM}},g)$ plane, where $m_{\textrm{DM}}\equiv m_\chi\simeq m_\phi$ and one assumes $m_N = 10~m_{\textrm{DM}}$, $y_L = 1$. ACT+Planck+BAO exclusion bounds obtained in this study are shown as a blue-shaded region, while the mean value of $\sigma_{\textrm{DM}-\nu}$ in our fit is obtained along the blue dashed line.}
    \label{fig:DMnuinteractionsgeneral}
\end{figure}

We indicate, in \autoref{fig:DMnuinteractionsgeneral},  the relic target line below which one predicts too large a  thermal DM abundance, while a correct value of $\Omega_\chi h^2$ can be obtained, e.g., in the asymmetric DM scenario~(\cite{Petraki:2013wwa,Zurek:2013wia}). In this case, the symmetric DM component can be efficiently annihilated away in the early Universe due to the $\chi\bar{\chi}\to \nu\bar{\nu}$ process. The remaining DM abundance driven by the initial asymmetry between $\chi$ and $\bar{\chi}$ can be higher than in the standard freeze-out. In this way one also avoids DM indirect detection bounds~(\cite{Arguelles:2019ouk}) as the number of DM antiparticles is depleted. In \autoref{fig:DMnuinteractionsgeneral}, we show with a blue-shaded region cosmological constraints on DM-neutrino interaction cross section that we obtain based on ACT+Planck+BAO data. We also present  colored lines with fixed values of the $u_{\textrm{urDM}}$ parameter between  $-3$ and $-8$, as well as with a blue dashed line the mean value of this parameter from our fit. For comparison, a light red-shaded region is shown, inside which Lyman-$\alpha$ observations can be better explained assuming non-negligible DM-neutrino interactions ($1\sigma$)~(\cite{Hooper:2021rjc}). The DM-neutrino interaction strength obtained this way lies remarkably close to the mean value of $\sigma_{\textrm{DM}-\nu}$ obtained in this work. Future cosmological data and Lyman-$\alpha$ observations will  constrain further the allowed region in the parameter space of this model. In \autoref{fig:DMnuinteractionsgeneral}, we also illustrate expected sensitivity of the Dark Energy Spectroscopic Instrument (DESI) to probe $\nu$DM interaction strength following Ref.~(\cite{Escudero:2015yka}) and a (optimistic) future bound on the $U_{\tau 4}$ mixing angle from the Belle-II experiment where larger couplings would be excluded~(\cite{Kobach:2014hea}).

\section{Conclusions}

\label{sec:discussion}}

In this work, we have analyzed the effects of the interaction between dark matter and neutrinos, assuming a temperature-independent interaction cross-section. Considering small-scale CMB data from the Atacama Cosmology Telescope, we find a preference for a non-zero interaction strength. This result remains consistent when combining observations from the two most accurate CMB experiments to date (Planck and ACT) and including astrophysical measurements of Baryon Acoustic Oscillations and Redshift Space Distortions. We have also indicated how scenarios involving a sterile neutrino portal between dark matter and the SM could accommodate such a coupling.

In order to validate the robustness of our findings, we have conducted a significant number of additional tests, all of which have confirmed this preference for a non-zero interaction. Specifically, we have observed the same preference when including or excluding BAO, and when varying or fixing the effective number of relativistic particles ($N_{\rm eff}$) in the cosmological model. Moreover, we have found that a similar preference emerges even when considering a temperature-dependent cross-section $\sigma_{\nu\textrm{DM}} \propto T^2$, indicating that this is not an artifact of assumptions made in the parameterization of the interaction~(\cite{Brax:2023tvn}).

To gain a better understanding of our results, we have thoroughly examined the data provided by both experiments and verified that the peak in the distribution of the interaction strength is associated with a genuine reduction of the $\chi^2$ of the fit. We have conducted a Bayesian model comparison to assess the plausibility of both interacting and non-interacting models in explaining the current observations. We found that while both models are plausible, the interacting case is often favored over the non-interacting one with moderate preference. We will present the results of all the additional tests in a separate work~(\cite{Brax:2023tvn}).

Finally, it is  important to note that the interaction strength value obtained from our analysis ($\log_{10}u_{\nu \textrm{DM}}=-5.20^{+1.2}_{-0.74}$) is consistent with the result obtained in Ref.~(\cite{Hooper:2021rjc}) from Lyman-$\alpha$ probes. The latter found a significant preference for an interaction strength ($\log_{10}u_{\nu \textrm{DM}}=-5.42^{+0.17}_{-0.08}$) approximately 3$\sigma$ away from zero when considering Lyman-$\alpha$ data. This effect is attributed to the  additional tilt in the Lyman-$\alpha$ flux power spectrum which affects small scales and leads to an  improved  fit compared to the $\Lambda$CDM model.  The remarkable correspondence between these two cosmological probes provides further hints of possible departures from the standard cosmological scenario. Interactions between DM and neutrinos can also affect the small-scale structure of the Universe and have been proposed to address some of the persisting problems of $\Lambda$CDM, e.g. the missing satellite issue, see~(\cite{Boehm:2014vja,Bertoni:2014mva,Schewtschenko:2015rno}). We leave detailed analyses of the interplay between these effects for future studies.
 
Our result will be testable and better bounds will be obtained with the next generation of CMB experiments, such as~(\cite{Abazajian:2019eic,SimonsObservatory:2018koc,NASAPICO:2019thw,CMB-HD:2022bsz}), see also Ref.~(\cite{Escudero:2015yka}) for expected sensitivity of DESI reaching up to $\log_{10}u_{\nu \textrm{DM}}\simeq -5.43$. Future surveys sensitive to high CMB multipoles will open a new window for probing dark matter couplings to neutrinos.

%%%%%%%%%%%%%%%%%%%%%%%%%%%%%%%%%%%%%%%%%%%%%%%%%%
\vspace{0.5cm}
\section*{Data Availability}

All the data used are explained in the text and are publicly available.
%The inclusion of a Data Availability Statement is a requirement for articles published in MNRAS. Data Availability Statements provide a standardised format for readers to understand the availability of data underlying the research results described in the article. The statement may refer to original data generated in the course of the study or to third-party data analysed in the article. The statement should describe and provide means of access, where possible, by linking to the data or providing the required accession numbers for the relevant databases or DOIs.

\section*{Acknowledgements}
ST would like to thank Brian Batell for useful discussions. CvdB is supported (in part) by the Lancaster–Manchester–Sheffield Consortium for Fundamental Physics under STFC grant: ST/T001038/1. EDV is supported by a Royal Society Dorothy Hodgkin Research Fellowship. ST is also supported by the grant ``AstroCeNT: Particle Astrophysics Science and Technology Centre" carried out within the International Research Agendas programme of the Foundation for Polish Science financed by the European Union under the European Regional Development Fund. ST is also supported in part by the National Science Centre, Poland, research grant No. 2021/42/E/ST2/00031. ST acknowledges the support of the Institut Pascal at Université Paris-Saclay during the Paris-Saclay Astroparticle Symposium 2022, with the support of the P2IO Laboratory of Excellence (program “Investissements d’avenir” ANR-11-IDEX-0003-01 Paris-Saclay and ANR-10-LABX-0038), the P2I axis of the Graduate School Physics of Université Paris-Saclay, as well as IJCLab, CEA, IPhT, APPEC, the IN2P3 master projet UCMN and ANR-11-IDEX-0003-01 Paris-Saclay and ANR-10-LABX-0038. This article is based upon work from COST Action CA21136 Addressing observational tensions in cosmology with systematics and fundamental physics (CosmoVerse) supported by COST (European Cooperation in Science and Technology). We acknowledge IT Services at The University of Sheffield for the provision of services for High Performance Computing.



\bibliographystyle{mnras}
\bibliography{main} % if your bibtex file is called example.bib

%\begin{thebibliography}{99}
%\end{thebibliography}

%%%%%%%%%%%%%%%%%%%%%%%%%%%%%%%%%%%%%%%%%%%%%%%%%%

%%%%%%%%%%%%%%%%% APPENDICES %%%%%%%%%%%%%%%%%%%%%

%\appendix



%%%%%%%%%%%%%%%%%%%%%%%%%%%%%%%%%%%%%%%%%%%%%%%%%%

% Don't change these lines
\bsp	% typesetting comment
\label{lastpage}
\end{document}